\documentclass[fleqn,twocolumn,10pt]{wlscirep}
\usepackage[utf8]{inputenc}
\usepackage[T1]{fontenc}
\usepackage{amsmath,amssymb,amsthm,geometry,mathtools}
\usepackage{enumitem,cleveref,subfigure,caption}
\usepackage{graphicx,booktabs,cuted,xcolor,stfloats}
\usepackage{url,lipsum,dashrule,indentfirst}

\newtheorem{theorem}{\textbf{Theorem}}

\newtheorem{lemma}{\bf Lemma}
\theoremstyle{definition}

\theoremstyle{remark}

\title{Meritocracy versus Matthew-effect: Two underlying network formation mechanisms of online social platforms }

\author[1]{Yuchen Xu}
\author[1,*]{Wenjun Mei}
\author[2]{Ge Chen}
\author[3]{Linyuan Lü}
\affil[1]{School of Advanced Manufacturing and Robotics, Peking University, Beijing, China}
\affil[2]{Academy of Mathematics and Systems Science, Chinese Academy of Sciences, Beijing, China}
\affil[3]{University of Science and Technology of China, Hefei, China}
\affil[*]{mei@pku.edu.cn}

\begin{abstract}
With the rapid development of the internet industry, online social networks have come to play an increasingly significant role in everyday life. In recent years, content-based emerging platforms such as TikTok, Instagram, and Bilibili have diverged fundamentally in their underlying logic from traditional connection-based social platforms like Facebook and LinkedIn. Empirical data on follower counts and follower-count-based rankings reveal that the distribution of social power varies significantly across different types of platforms, with content-based platforms exhibiting notably greater inequality.
Here we propose two fundamental network formation mechanisms: a meritocracy-based model and a Matthew-effect-based model, designed to capture the formation logic underlying traditional and emerging social networks, respectively. Through theoretical and numerical analysis, we demonstrate that both models replicate salient statistical features of social networks including scale-free and small-world property, while also closely match empirical patterns on the relationship between in-degrees and in-degree rankings, thereby capturing the distinctive distributions of social power in respective platforms. Moreover, networks such as academic collaboration networks, where the distribution of social power usually lies between that of traditional and emerging platorms, can be interpreted through a hybrid of the two proposed mechanisms. Deconstructing the formation mechanisms of online social networks offers valuable insights into the evolution of the content ecosystems and the behavioral patterns of content creators on online social platforms.
\end{abstract}
\begin{document}

\flushbottom
\maketitle
%
%
\thispagestyle{empty}

\section*{Introduction}


During the past decade, the widespread adoption of smartphones, computers and other devices has fostered the booming development of online social networks. 
With the emergence of an increasing variety of platforms and the continuous growth of user populations, these platforms have become deeply embedded in daily life, serving as important channels for information exchange, knowledge acquisition, and resource sharing \cite{bakshy2012role, gibbs2007social, bandiera2006social, burkhardt1990changing}.
Nevertheless, despite all being classified as online social platforms, substantial differences in both structure and function can be observed across platforms, as both how users interact with social platforms and the purposes behind their engagement have evolved \cite{10.1145/2517040, 10.1145/1519065.1519089, 10.1108/OIR-06-2020-0258, 6289250}.
On earlier platforms such as Facebook, users' engagement primarily took the form of sharing personal daily updates. Today, with the enrichment of digital content formats and the enhancement in information diffusion/acceleration of information flows, platforms like TikTok now allow users to create and share diverse forms of content within specific categories or topics. These changes reflect a fundamental transition in the underlying logic of social media.


Building on these observations, at a conceptual level, online social platforms can be broadly grouped into two paradigms based on how user relationships are structured and sustained. On traditional online social networks such as Facebook and LinkedIn, users’ popularity serves as the primary factor influencing the acquisition of new followers in addition to their offline acquaintances, and the Matthew effect plays an important role during the link formation process. 
In contrast to such kinds of connection-based / popularity-oriented networks, those arising online platforms such as Tiktok or Twitch can be understood as content-based /quality-oriented networks. On these platforms, users attract the attention of strangers with whom they have no prior offline relationships by producing high-quality content within specific categories. 
Additionally, some platforms exhibit intermediate characteristics, where the way users form connections is influenced by both popularity and quality, such as scientific collaboration networks.
These developments have sparked interest in whether and how such fundamental transition in underlying mechanisms could exert notable influence on the structural characteristics they exhibit. 

The past few decades have witnessed a flourishing of research on how social networks form into certain observed structures \cite{wasserman1994social, freeman2004development, doreian2012social, li2021review}, among the most common are random graph models focusing on macroscopic structural statistics such as degree distribution, diameter, and clustering coefficient. One form of the Matthew effect is modeled by the classic preferential attachment framework proposed by Barábasi and Albert \cite{barabasi1999emergence} along with various extensions and applications \cite{papadopoulosPopularitySimilarityGrowing2012, capocci2006preferential, kunegis2013preferential}. Other well-known models include small-world network model proposed by Erdös and Rényi \cite{watts1998collective, PhysRevE.60.7332}, strategic network formation models widely used in economics \cite{jackson2008social, slikker2012social, skyrms2009dynamic}, etc. These earlier models are primarily designed for connection-based networks and generally do not incorporate content quality as a fundamental factor in the formation process of social networks. In a recent study \cite{pagan2021meritocratic}, Pagan et al. proposed a quality-based model in which individuals tend to form links towards higher-quality users. Theoretical results predict that individuals’ expected in-degrees follow the well-known Zipf’s law with respect to their ranking. However, further empirical analysis reveals evident discrepancies between the model’s predicted indegree–ranking relationship and patterns observed in real-world data, suggesting that content-based platforms requires further modeling efforts.

From another perspective, besides having been given an interpretation in the purview of network science, the Matthew effect is also widely regarded as a classical explanation and a significant driving force behind inequality in the distribution of social power \cite{perc_matthew_2014, bothner_when_2010, 799a3f11-1163-3207-808d-db1b878f6b31, janicka_polarized_nodate}. Social power is a fundamental concept in sociology, widely used in characterizing an individual's influence among a group or community. Analyzing its distribution is crucial for understanding the features of social interactions and the formation of social structures \cite{bierstedt_analysis_1950, bandiera2006social, burkhardt1990changing, miller_social_2008, kellerman_political_1986, Bramson_Grim_Singer_Berger_Sack_Fisher_Flocken_Holman_2017}. Within the context of social networks, social power is often interpreted as node degree or other kinds of centrality measure, with its distribution pattern commonly measured by indicators such as the Gini coefficient \cite{barabasi1999emergence, jackson2008social, wasserman1994social}. Traditionally, the power-law distribution resulting from preferential attachment networks is often regarded as a classic example of unequal social power distribution.
However, our recent observations reveal that emerging social platforms, grounded in user-generated content (UGC) and driven primarily by quality rather than popularity, exhibit even greater inequality in social power distribution compared to traditional platforms (See Figure \ref{fig:data}). This phenomenon warrants close attention because, although UGC ostensibly democratizes social influence by giving users more equal opportunities in content generation and expression, it paradoxically leads to increased social inequality. Therefore, providing a mechanistic explanation for this surprising outcome is of great significance.

In this paper, we propose two parsimonious network formation mechanisms: the meritocracy-based mechanism and the Matthew-effect-based mechanism, as explanatory models for the underlying formation process of online social platforms. For content-based platforms, we introduce a meritocratic principle to control the link formation process; while for connection-based platforms, we adopt the classic preferential attachment mechanism to characterize network formation process. In both models, we incorporate a limited-attention constraint by imposing a uniform upper bound on the out-degree of each node, which captures the inherent limitation of users' cognitive resources on online platforms \cite{dukas_causes_2004, lerman_limited_2013, weng_competition_2012, wang_social_2019}. This assumption is not only intuitively reasonable but also empirically supported by real-world data.

Through analytical and numerical analysis, we find that both proposed models exhibit statistical characteristics commonly observed in traditional social networks, such as the scale-free and small-world properties.
Moreover, we analyze the relationship between nodes' expected in-degree and their ranking (in descending order of expected in-degree), as a measure to characterize the distribution of social power in different types of online social networks. We provide rigorous theoretical analysis for the meritocracy-based model and derive analytical approximations of expected in-degree for both models. 
Together with simulation results, we observe that in the meritocracy-based model, top-ranked nodes possess significantly higher in-degrees than others, and the competition among themselves is also particularly intense, resulting in a higher Gini coefficient of expected in-degrees. By comparison, in the Matthew-effect-based model, the corresponding Gini coefficient is lower as in-degree decreases more gradually with ranking. 
This discrepancy lies beyond the discriminative capacity of conventional network statistical measures, and the conclusions derived from theoretical models closely align with patterns observed in empirical data. This supports a seemingly surprising finding that shifting from the Matthew-effect-based to a meritocracy-based link formation mechanism may lead to an even more unequal distribution of social power on emerging social platforms. In addition, to explore the intermediate cases, we construct a hybrid model by combining the meritocracy-based and Matthew-effect-based mechanisms in varying proportions, such that each node forms links based on a probabilistic choice between the two mechanisms. We find that as the mixing ratio changes monotonically, the indegree–ranking relationship changes in a correspondingly monotonic way, capturing a continuum between different types of network. This hybrid approach successfully captures the social power distribution patterns of real-world networks that appear to lie between purely content-based and purely connection-based ones. These results suggest that the meritocracy-based and Matthew-effect-based mechanisms together constitute two fundamental dimensions of the underlying formation mechanism of online social networks.

\section*{Results}

\subsection*{Data analysis}

\begin{figure*}[ht]
   \centering
   \includegraphics[width=\textwidth]{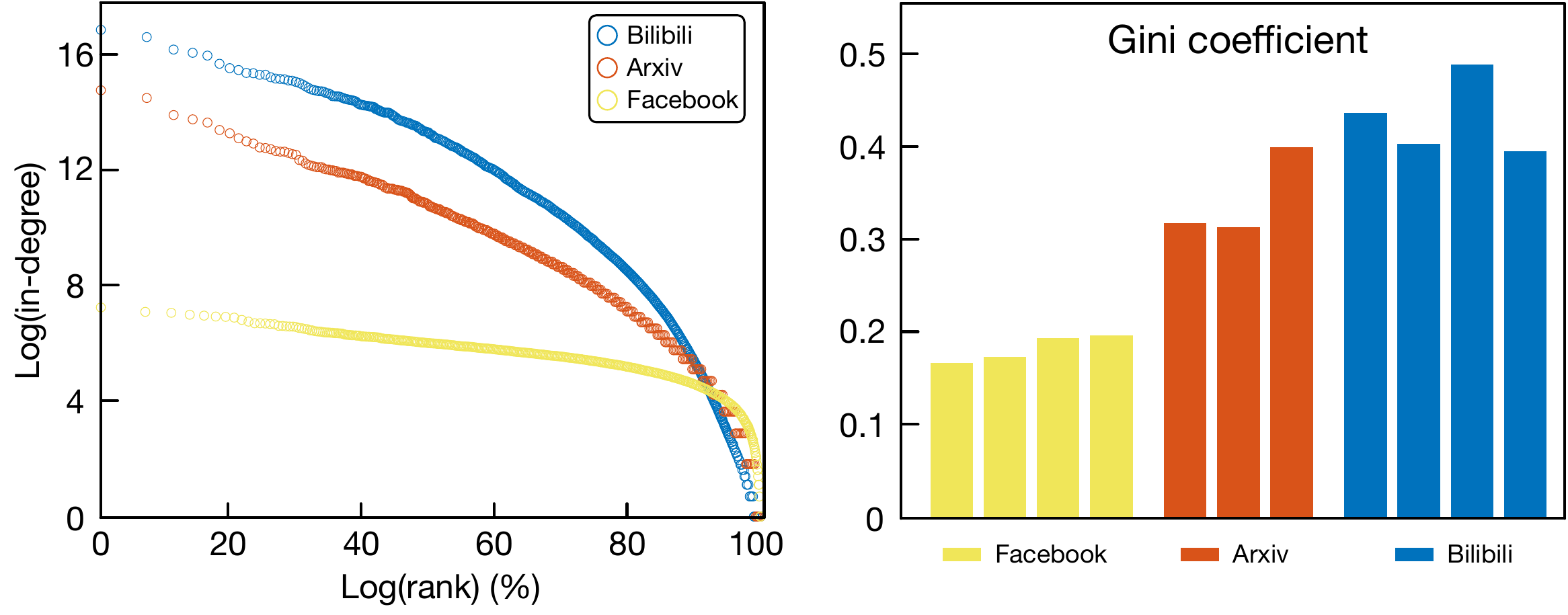}
   \caption{\textbf{| Data analysis of real-world online social platforms.} \textbf{A} Relationship between users' follower counts and their follower-count-based ranking on connection-based networks (e.g., Facebook Indiana university), content-based networks (e.g. Bilibili Science \& Education category), and mixed-mechanism networks (e.g. ArXiv Condensed Matter collaboration network) under log-log coordinates. Network link density is normalized for comparability, and consistent results have been observed across different categories and communities of various online platforms. \textbf{B} Gini coefficients of follower counts across different types of platforms: Facebook (Indiana, Georgetown, Berkeley, Michigan), Arxiv collaboration networks (Condensed Matter, Astro Physics, High Energy Physics), and Bilibili (Science \& Education, Food Reviews, Daily Life, Anime Commentary).}
   \label{fig:data}
\end{figure*}

To investigate the patterns of social power distribution across different types of online platforms, we collect real-world data and numerically analyze the relationship between users' follower counts and their follower-count-based ranking under log-log coordinates \cite{nr-aaai15,snapnets}. 
Empirical evidence in Figure \ref{fig:data} suggests that the transition from connection-based networks to content-based networks is accompanied by a monotonic shift in the relationship between users' follower counts and their corresponding rankings. 
While most social platforms share a general pattern in which follower counts initially scale linearly with ranking then declines more rapidly at lower rankings, as networks shift from connection-based to mixed-mechanism and eventually to content-based ones, follower counts among top-ranked users show a consistent increase, whereas those of mid-tier and lower-tier users decline, indicating a growing concentration of social power among top influencers. Although the intersection of indegree-ranking curves appears in relatively bottom areas, the compressive nature of logarithmic scaling means that the share of top-ranked nodes deriving such advantage is comparatively modest. The increasing inequality in social power distribution is further supported by the monotonic trend in the Gini coefficients of follower counts across various categories or communities of different platforms.
Such disparity in social power distribution reflects a dimension of social networks that extends beyond traditional structural properties such as power-law degree distributions, short network diameters or high clustering coefficients. Hence, providing a mechanistic explanation of this phenomenon could shed light on the evolution of the content ecosystem and user landscape on online social platforms.

We argue that these differences arise from the underlying mechanisms by which the networks are formed. Specifically, connection-based networks are shaped by the Matthew effect, where individuals with higher existing follower counts are more likely to attract new connections. In contrast, content-based networks develop under the meritocratic principle \cite{pagan2021meritocratic}, in which an individual's follower count is directly influenced by the quality of UGC. Scientific collaboration networks, meanwhile, lie somewhere in between: researchers tend to consider both the prominence of a potential collaborator's existing connections and the quality of his academic output. 
To more precisely characterize the two underlying mechanisms and to further understand the social power distribution patterns across different types of networks, we formalize each mechanism by proposing a theoretical network formation model.

\subsection*{Model setup}

\begin{figure*}[ht]
   \centering
   \includegraphics[width=\textwidth]{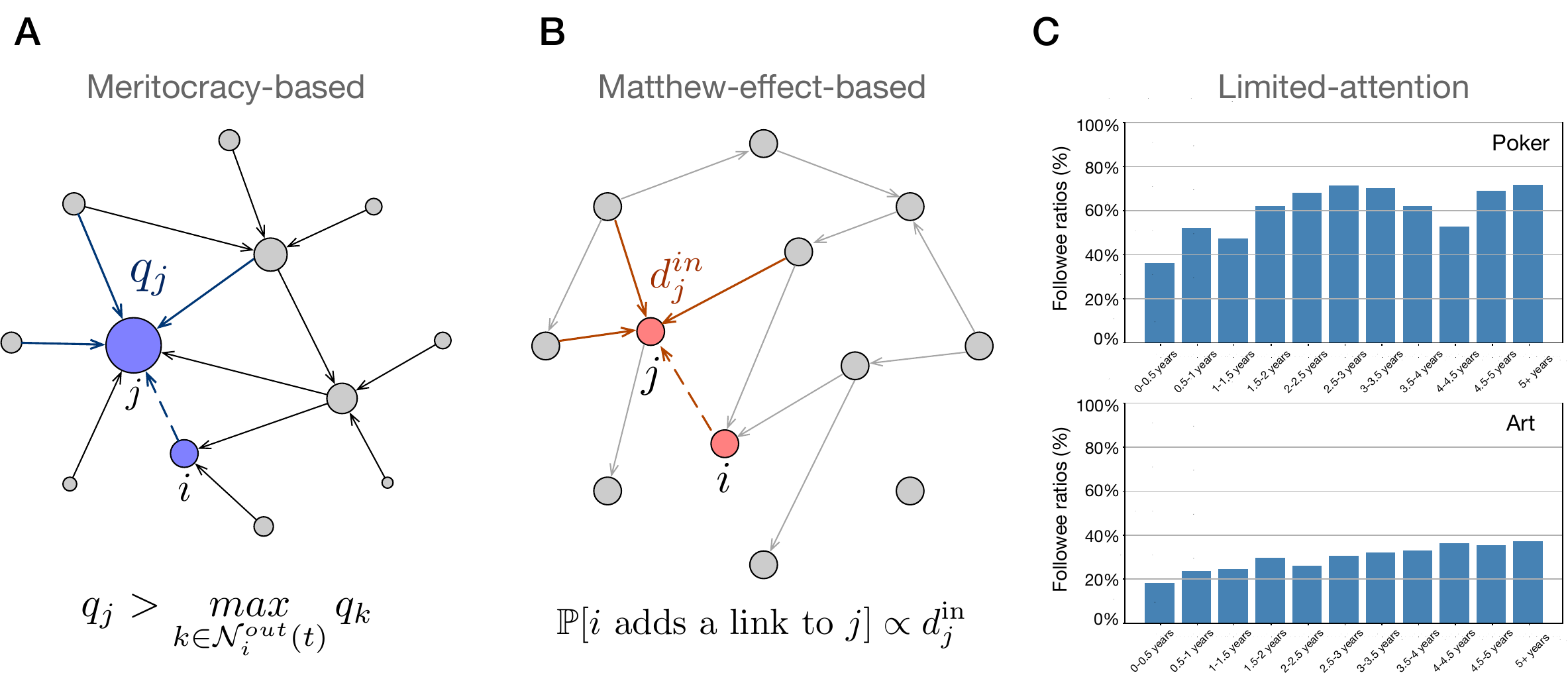}
   \caption{\textbf{| Network formation mechanism of the proposed models.} Illustration of network formation mechanism in \textbf{A} Matthew-effect-based model and \textbf{B} meritocracy-based model, where the former is primarily influenced by in-degree $d_j^{\text{in}}$, while the latter is determined by content quality $q_j$. \textbf{C} Empirical evidence of the limited-attention assumption on Twitch: variation in the proportion of followed streamers in the poker and art categories with respect to user account age.}
   \label{fig:model}
\end{figure*}

Our study posits that the formation of online social networks is fundamentally shaped by two key mechanisms. Links generated under the meritocracy-based mechanism are driven by the intrinsic quality of nodes, whereas those formed under the Matthew-effect-based mechanism are influenced by the existing in-degree of nodes. The two mechanisms represent distinct principles in network formation process and each can be formally characterized through mathematical modeling, as illustrated in Figure \ref{fig:model}. 

We first introduce the meritocracy-based model. Consider an unweighted directed network among $N \geq 2$ nodes. Define a binary variable $a_{ij}$, where $a_{ij} = 1$ indicates that there is a directed link from node $i$ to node $j$, i.e., $i$ is a follower of $j$ on the social network, and $a_{ij} = 0$ otherwise. We assume that there is no self-loop in the network, i.e., $a_{ii}=0$, for any $i$. For each node $i\in\{1,2,...,N\}$, we assign an attribute $q_i \in \mathbb{R}$ which represents the quality of content generated by $i$. For the convenience and simplicity of writing, we assume that $q_1>q_2>...>q_N$, which can always be obtained by re-indexing when qualities are different from each other. For situations where $q_1 \geq q_2 \geq...\geq q_N$, similar results can be obtained analogously. 

The formation process starts with an empty network. At each time step $t \in \{1,2,...\}$, two different nodes $i,j$ are uniformly randomly selected. Under the meritocratic principle \cite{pagan2021meritocratic}, node $i$ will follow node $j$ if and only if the quality of $j$ is higher than any of $i$'s current followees. Therefore, what really matters is the ordering of qualities rather than their magnitudes. Moreover, we introduce a limited-attention assumption, namely that the out-degree of any node $i$ should never exceed a given threshold. This assumption is supported by empirical evidence presented in Figure \ref{fig:model}C. Therefore, the meritocracy-based network is updated according to the following rule: 

\begin{equation*}
	a_{ij}(t+1) = 
	\begin{cases}
	\begin{aligned}
		&1\quad\enspace\ ,\ \text{if}\ q_j > \mathop{max}_{k\in \mathcal{N}_{i}^{\text{out}}(t)} q_k\ \text{and}\ d_i^{\text{out}}(t)<M \\
		&a_{ij}(t),\ \text{otherwise}
		\end{aligned}\ ,
	\end{cases}
\end{equation*}
where $\mathcal{N}_{i}^{\text{out}}(t)$ represents the set of all followees of node $i$ at time step $t$, and $M \in \mathbb{Z}^{+}$ is the prescribed out-degree constraint. We assume that $\mathop{max}_{k\in \emptyset} q_k=-\infty$.

We claim that the network formation process converges almost surely. For any node $i \in \{1,2,...,N\}$, there are only two conditions under which node $i$ will stop building new out-links at time step $t$ (i.e. reach an equilibrium state):

\begin{enumerate}[label=\arabic*)]

\item Node $i$ has already been linked to the node with highest quality, i.e., 
$a_{i2}(t)=1\enspace \text{if}\enspace i=1;\enspace a_{i1}(t)=1\enspace \text{if}\enspace i \neq 1$;

\item Node $i$'s out-degree has reached the upper bound $M$, i.e., $d_i^{\text{out}}(t)=M$.

\end{enumerate}

Meanwhile, it is worth mentioning that $a_{ij}(t+1)=a_{ij}(t)=1$, if $j \in \mathcal{N}_{i}^{\text{out}}(t)$. Namely, existing links will never be removed, indicating that the out-degree of any node in the network increases monotonically with respect to time $t$ and is subject to the same threshold $M$. Therefore, the network reaches an equilibrium almost surely (See Methods).

For the Matthew-effect-based model, its link formation process can be characterized by the classic preferential attachment mechanism. Similarly, consider an unweighted directed network among $N \geq 2$ nodes. At the initial state, each node owns a virtual self-link which does not count in its in-degree and there is no links between different nodes. At each time step $t \in \{1,2,...\}$, randomly select a node $i$ with uniform probability and add a directed edge from $i$ to $j$, where the probability of node $j$ been chosen is proportional to its current in-degree
\[
    \mathbb{P}[\text{node }j\text{ is chosen at time }t] \propto d_j^{\text{in}}(t)
\]
We also incorporate the limited-attention assumption so that the out-degree of any node $i$ should never exceed a given threshold $M \in \mathbb{Z}^+$. Therefore, the network formation process will end at $t=M\cdot N$ when every node has precisely $M$ out-links.

\subsection*{Typical patterns of social networks}

\begin{figure*}[ht]
   \centering
   \includegraphics[width=\textwidth]{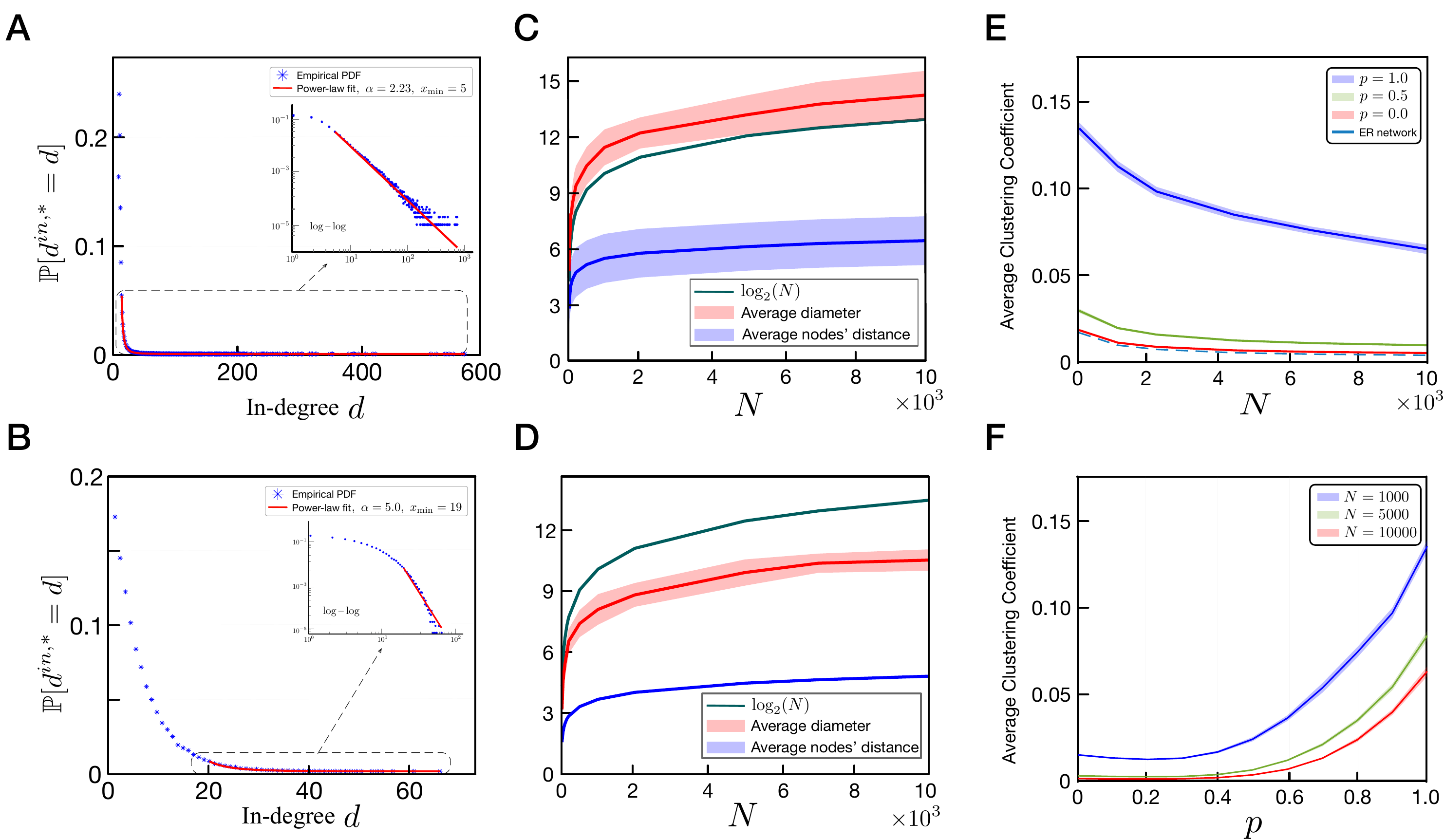}
   \caption{\textbf{| Verification of scale-free and small-world property.} \textbf{A,B} Power-law in-degree distribution verification of the meritocracy-based model (\textbf{A}) and the Matthew-effect-based model (\textbf{B}), using a network size of $N = 10,000$. \textbf{C,D} Network diameter and average path length calculation of the meritocracy-based model (\textbf{C}) and the Matthew-effect-based model (\textbf{D}), together with a $\log_2(N)$ benchmark curve. \textbf{E} Average clustering coefficient calculation of the meritocracy-based model ($p=1$), the Matthew-effect-based model ($p=0$) and a hybrid model combining the two mechanisms with equal probability ($p=0.5$). The average clustering coefficient of a directed Erdős–Rényi model with same network density is provided as a reference. \textbf{F} Average clustering coefficient under different mixing ratio $p$. Simulation experiments were conducted with $100$ runs for each model under an in-degree constraint of $M = 5$. All the results remain consistent for different values of $N$ and $M$ (See Supplementary).}
   \label{fig:SWandSF}
\end{figure*}

Despite representing two fundamental dimensions of the network formation mechanisms underlying online social platforms, the two proposed models still share certain commonalities in replicating salient statistical features commonly observed in social networks, such as power-law in-degree distribution and low network diameter and average path length. Numerical verification for both models is given in Figure \ref{fig:SWandSF}.

Regarding the scale-free property for in-degree distribution, it can be concluded from Figure \ref{fig:SWandSF}A,B that the meritocracy-based model exhibits a power-law distribution over the majority of in-degree values, deviating only in the region of very low in-degrees \cite{clauset2009power}. In comparison, the Matthew-effect-based model supports a power-law distribution across a narrower range of in-degrees, though the conclusion still holds for large $d$, capturing the long-tail behavior that characterizes power-law distributions. Therefore, we conclude that at the equilibrium state, the in-degree distribution in both models broadly follows a power-law distribution. Further analytical results are provided to support this conclusion (see Methods).

Complementing the analysis of scale-free behavior, we also verify the small-world property through numerical analysis, which is popularly known as six degrees of separation \cite{watts1998collective, milgram_small-world_1967, kochen1989small, Guare2016}. We first clarify the definitions of diameter, average path length, and clustering coefficient in the proposed models. Since diameter is originally defined in connected graphs \cite{jackson2008social}, while the corresponding graph of both models might not satisfy this condition, we redefine diameter as the maximum length of shortest directed paths between all connected pairs of nodes.
Since we care more about those high-ranking influencers who owns the vast majority of in-links at the equilibrium state, such a modification is reasonable and aligns with our original motivation for analyzing network diameter. Similarly, we only consider the connected pairs of nodes when computing the average nodes' distance. As for the clustering coefficient, define \( b_{ij} = a_{ij} + a_{ji} \) where \( A = [a_{ij}] \) is the adjacency matrix. Then the clustering coefficient of node \( i \) is given by
\[
C_i = \frac{ \frac{1}{2} \sum_{j \ne i} \sum_{k \ne i, j} b_{ij} b_{jk} b_{ki} }{ \left( \sum_{j \ne i} b_{ij} \right) \left( \sum_{j \ne i} b_{ij} - 1 \right) },
\]
which is a measure of the degree to which nodes in a digraph tend to cluster together \cite{jackson2008social, fagiolo2007clustering}.

Figure \ref{fig:SWandSF}C,D show that for both models, the network diameter has a growth rate similar to $\log_2N$ and the average nodes’ distance has a slower growth with similar trend, which match the empirical observations on the small-world property of real-world networks \cite{watts1998collective}. It can be further confirmed that this conclusion holds for different $M$. Therefore, in summary we can draw the conclusion that both the meritocracy-based model and the Matthew-effect-based model capture the scale-free property as well as the small-world property.
As for the clustering coefficient, Figure \ref{fig:SWandSF}E indicates that compared to a directed Erdős–Rényi(ER) graph model\cite{erdHos1960evolution, lima2008majority} with the same link density, the meritocracy-based model exhibits a significantly higher clustering coefficient, whereas the Matthew-effect-based model yields a clustering coefficient that is roughly comparable to that of the ER graph. Furthermore, when a probabilistic combination of the two mechanisms is introduced, we observe that the clustering coefficient generally increases with the mixing parameter $p$, where higher values of $p$ indicate that a greater proportion of links are formed under the meritocracy-based mechanism.

\subsection*{Social power distribution analysis}

While the two theoretical models may offer a plausible representation of the mechanisms underlying different types of online social platforms, a more convincing validation lies in their ability to reproduce the observed patterns of social power distributions, namely, the relationship between nodes’ expected in-degree and their ranking. This calls for a theoretical analysis of the expected in-degree at the equilibrium state for both models.

In the meritocracy-based model, node $i$’s indegree-based ranking naturally aligns with its intrinsic quality $q_i$. However, as previously noted, nodes may stop creating new out-links due to either the meritocratic principle or the out-degree constraint, and this uncertainty regarding the condition of reaching an equilibrium complicates the derivation of the in-degree distribution. Here we establish a recursive formula to facilitate the computation of nodes' expected in-degree in the meritocracy-based model.

Denote by $P(m,i)$ the total probability over all nodes other than $i$ of being linked to $i$ in their first $m$ out-links during the network formation process. If node $j$ has less than $m$ out-links at the equilibrium state (i.e. $d_j^{out}<m$), then the ``first $m$ out-links" refers to all out-links owned by node $j$. Through theoretical analysis we can derive the following recursive formula:
\begin{equation}\label{eq:P_recur}
	P(m,i-1)=P(m,i)+\frac{1}{i-1}P(m-1,i), \ 2 \leq i \leq N
\end{equation}
with boundary conditions $P(1,i)=P(m,N)=1,\  1 \leq i \leq N, 1 \leq m \leq M$ (see Supplementary for more details).
Since the out-links of different nodes are independent of each other, calculating the expected in-degree of node $i$ under out-degree constraint $M$ can be achieved by separately determining whether every other node $j$ follows $i$ and then summing them up. Therefore, the expected in-degree of any node $i$ can be precisely computed by the following explicit formula:
\begin{equation}\label{eq:explicit indegree}
\begin{aligned}
	\mathbb{E}[d_i^{in}] = P(M,i) =	\sum_{k=0}^{M-1} 
	\left ( 
	\sum_{A: A \subset \{i,i+1,...,N-1\},\ |A|=k}\  \prod_{j \in A} \frac{1}{j}
	\right ),
\end{aligned}
\end{equation}
where $\prod_{j \in \emptyset} \frac{1}{j} = 1$. For sufficiently large network size $N$, by using integrals to approximate sums, we can obtain an approximation (see methods):
\begin{equation}\label{eq:approximate indegree}
\begin{aligned}
	\mathbb{E}[d_i^{in}] \approx \sum_{m=0}^{M-1} \frac{1}{m!}(\log\frac{N}{i})^m,
	\ 1 \leq i \leq N, N \rightarrow \infty.
\end{aligned}
\end{equation}


We now turn to the analysis of the Matthew-effect-based model, in which an equilibrium state is reached only when every node reaches their maximum out-degree. However, since all nodes begin with equal standing in terms of in-degree, some preliminary preparations are required before adopting the mean-field approximation method to calculate the expected in-degree.

At each time step $t$, denote by $\mathbb{E}[d_i^{in}(t)]$ the expected in-degree of the $i^{th}$-ranked node (in terms of expected in-degree). Under the preferential attachment mechanism, with only one random link added to the network at $t=1$, we should have $\mathbb{E}[d_1^{in}(1)] = 1, \mathbb{E}[d_2^{in}(1)]=...=\mathbb{E}[d_N^{in}(1)] = 0$. By analogy, at time step $t=N$, we should have $\mathbb{E}[d_i^{in}(N)] \approx \frac{4N^2}{(N+i-1)(N+i)} - 1,\ i=1,2,...,N.$
With the initial conditions distinguishable among all nodes, by adopting the mean-field approximation method together with a continuous-time approximation of in-degree, we have
\[
    \frac{\mathrm{d}\mathbb{E}[d_i^{in}(t)]}{\mathrm{d}t} = \frac{\mathbb{E}[d_i^{in}(t)]}{N+t}.
\]
At the equilibrium state $t = M \cdot N$, the expected in-degree of the $i^{th}$-ranked node can thus be approximated by

\begin{equation}\label{eq:Matthew indegree}
    \mathbb{E}[d_i^{in}] \approx \frac{2(M+1)N^2}{(N+i-1)(N+i)} - 1,\ i = 1,2,...,N.
\end{equation}

\begin{figure*}[ht]
   \centering
   \includegraphics[width=\textwidth]{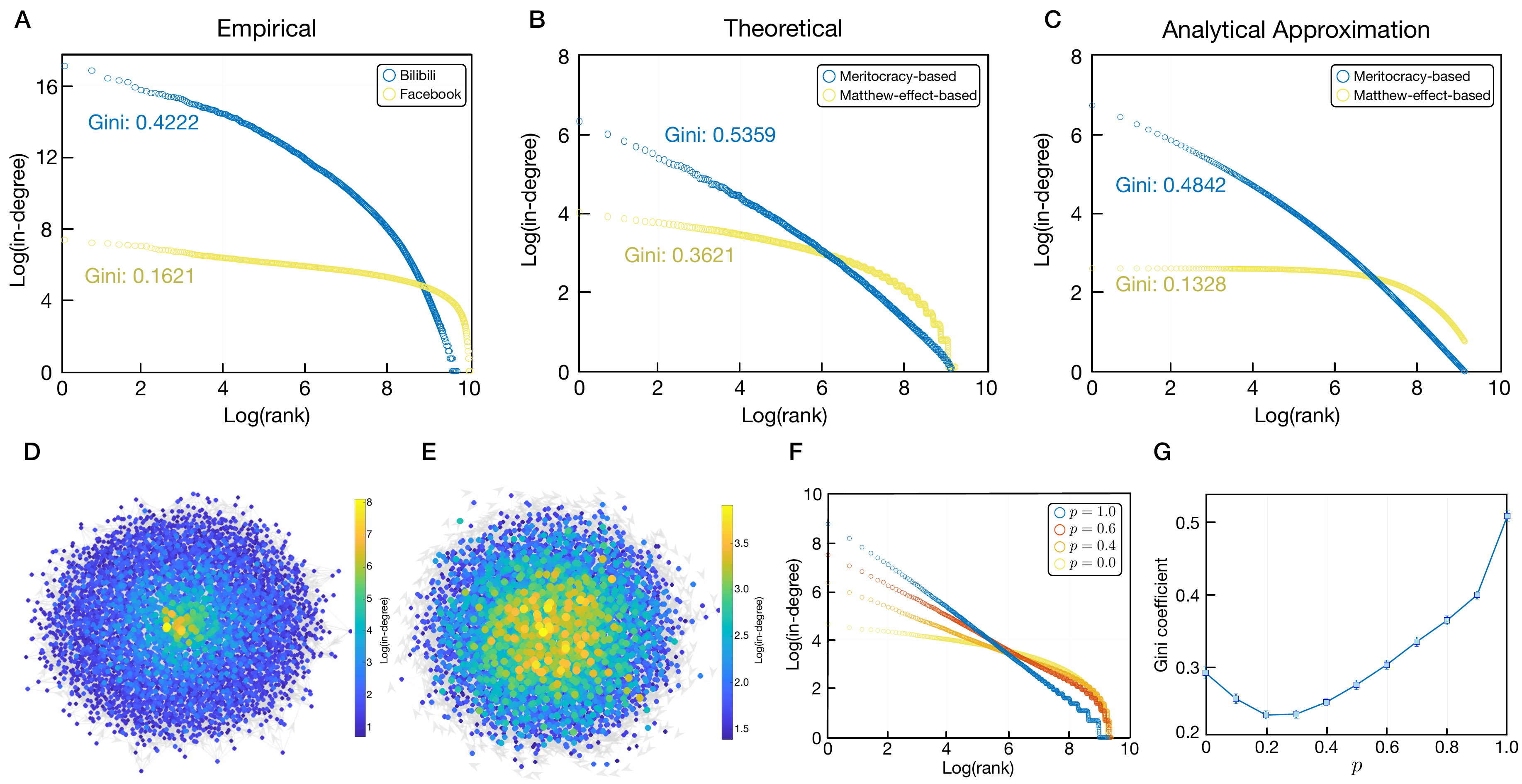} 
   \caption{\textbf{| Variations in social power distribution across online social networks.} \textbf{A-C} Comparison of social power distributions in (\textbf{A}) content-based platforms vs. connection-based platforms based on real-world data, and (\textbf{B,C}) the meritocracy-based model vs. the Matthew-effect-based model, based on simulation results and analytical approximation results of the proposed theoretical models.  \textbf{D,E} Visualization of in-degrees for the top 5,000 nodes in the meritocracy-based model and Matthew-effect-based model. \textbf{F} Social power distributions resulting from different proportions of the two network formation mechanisms, with $p = 0$ indicating a fully Matthew-effect-based process and $p = 1$ indicating a fully meritocracy-based process. \textbf{G} Variation in the Gini coefficient with respect to the mixing proportion $p$ between the two network formation mechanisms. All conclusions based on the theoretical models are illustrated under a network size of $N=10000$ and an out-degree constraint of $M=5$, with similar patterns observed across different values of $M$.}
   \label{fig:Results}
\end{figure*}

To further compare the distribution of social power in the two models, we introduce the Gini coefficient, which is a classical measure of inequality. By using the scale invariance of the Gini coefficient and verifying a single-crossing condition between the expected in-degree \eqref{eq:approximate indegree}, \eqref{eq:Matthew indegree} of the proposed models, we analytically conclude that the Gini coefficient of expected in-degrees for the meritocracy-based model is higher than that of the Matthew-effect-based model (see Supplementary), indicating that the distribution of social power in the former is relatively more unequal.

Moreover, building on the theoretical results above, we analyze the patterns of social power distribution in both models and compare them with the empirical characteristics observed in real-world data. Figure \ref{fig:Results} illustrates the relationship between nodes' expected in-degree and their ranking for the two proposed network formation models. All the theoretical results align closely with the empirical data, suggesting that our proposed models provide not only an intuitive interpretation, but also a theoretically and statistically grounded characterization of the underlying mechanisms behind different types of real-world social networks.

Moreover, both the results obtained from simulations and the results derived through analytical approximations \eqref{eq:approximate indegree}, \eqref{eq:Matthew indegree} consistently reveal a notable difference in the distribution of social power between the meritocracy-based model and the Matthew-effect-based model, mirroring the divergence observed across different types of social platforms. 
Specifically, under the same out-degree constraint $M$, the meritocracy-based model is characterized by top-ranked nodes having both high in-degree and considerable differentiation among themselves. This indicates that social power is primarily concentrated among top users, who exhibit significant disparities from mid-tier users and engage in intense competition among themselves. In contrast, the Matthew-effect-based model results in lower top-node in-degrees and a gentler decrease, suggesting a relatively more balanced distribution of social power. Consistent evidence is provided by the Gini coefficients computed in both theoretical and empirical results, which aligns with our analytical conclusions.


A further comparison of the two models reveals that, under a fixed value of out-degree constraint $M$, their indegree–ranking curves will intersect at only one single point under log-log coordinates. Nodes ranked above this point tend to gain more social power under the meritocracy-based mechanism, while those ranked below benefit more from the Matthew-effect-based mechanism. This indicates that the transition in online social platforms has led to a notable decline in social power among mid-tier users, as their influence is increasingly absorbed by top users on emerging content-based platforms. 
Therefore, compared to the Matthew-effect-based model, the meritocracy-based model elevates the head and suppresses the tail of the indegree-ranking curve, leading to a more unequal distribution of social power. This result indicates that a quality-oriented framework tends to foster the emergence of super influencers and intensifies competition among top users. 

Interestingly and somewhat surprisingly, although the Matthew effect is often considered as a major cause of inequality and social stratification, attempts to mitigate its influence by introducing content quality as the basis for building relationships may in fact exacerbate inequality in social power distribution on emerging social platforms. Notably, this outcome is determined solely by the relative ranking of users' content quality, regardless of its absolute value.

\subsection*{Cocktail model analysis}

Empirical evidence in Figure \ref{fig:data}B has shown that certain types of social networks, such as scientific collaboration networks, lie between connection-based and content-based ones in terms of social power distribution. Based on the analysis above, an intuitive idea is to consider whether a probabilistic combination of the meritocracy-based mechanism and Matthew-effect-based mechanism could give rise to intermediate patterns of social power distribution. Motivated by this, we introduce a probability parameter $p$ to control the mixing ratio between the two mechanisms and explore how the mixture probability influences the relationship between individuals' in-degree and ranking. 


Figure \ref{fig:Results}F presents the variation in the relationship between individuals' in-degree and ranking under different values of the mixing probability $p$, where $p=0$ represents a fully Matthew-effect-based model and $p=1$ indicates a fully meritocracy-based one. We observe that as the mixing probability $p$ varies, a hybrid model exhibits certain patterns of social power distribution that lie between the two ends. This aligns with the intuition that the formation of scientific collaboration networks is jointly shaped by individuals' content quality as well as connections. In addition, Figure \ref{fig:Results}G shows that as the mixing probability $p$ increases, the Gini coefficient of expected in-degree first undergoes a slight fall then increases dramatically. In general, for the majority of the parameter range, the variation in Gini coefficient is monotonic with respect to $p$ and aligns well with our expectation, which reveals a continuous and consistent increase in the inequality of social power distribution as the network formation mechanism shifts from the Matthew-effect-based one (underlying connection-based platforms) to the meritocracy-based one (underlying content-based platforms).

The anomalous trend observed at low values of $p$ can also be analytically explained. With a slight incorporation of meritocratic principle, the network formation process remains primarily dominated by preferential attachment. However, compared with the case of $p=0$, top-quality nodes are now able to occasionally gain additional in-degrees through the meritocratic principle, thus resulting in a greater number of individuals with high in-degree. This observation is further supported by the overlap between nodes with high in-degree and those with top quality (See Supplementary Figure 1). Given that the variation in Gini coefficient is relatively minor when $p$ takes a small value, while the subsequent changes become substantially more dramatic as $p$ increases, the overall trend supports the conclusion that a hybrid network formation model could be used to characterize real-world social networks that lie between content-based ones and connection-based ones. This justifies the idea of positioning the meritocracy-based and Matthew-effect-based mechanism as two representative dimensions of the underlying mechanisms of online social platforms. 

\section*{Discussion}

With the rapid development of the Internet industry and mobile devices,  online social networks have come to play an important role in facilitating information dissemination, opinion exchange, and knowledge acquisition. The evolution of media and formats for information sharing on these platforms has significantly expanded the range of user behaviors, enabling users to engage in a broader set of actions than before. This marks a shift from simply sharing personal updates to actively creating content in richer formats such as text, images, and videos. This evolution has given rise to a new type of content-based social platforms, which stand in contrast to traditional connection-based platforms. On these platforms, users now attract followers to gain social power primarily through producing high-quality content rather than cultivating interpersonal connections. And this shift in user behaviors has reshaped the structure of the corresponding online social networks. One salient feature revealed by our empirical analysis is the markedly more unequal distribution of social power on arising platforms, where in-degree resources are heavily concentrated among top users.

In this paper, we propose two network formation models, namely the meritocracy-based model and the Matthew-effect-based model, to explain the underlying mechanisms of different types of online social platforms. Through both theoretical and numerical analysis, we demonstrate that, by combining a parsimonious link formation principle with a limited-attention assumption, both models not only satisfy scale-free and small-world property, but also capture the distinct social power distribution patterns of content-based and connection-based networks, respectively. These findings suggest that the proposed models provide a mechanistic explanation and interpretation for the formation of online social networks, and lead to the conclusion that replacing the Matthew effect with a meritocratic principle in forming user connections may further exacerbate inequality in the distribution of social power. 
Furthermore, by introducing a convex combination of the two mechanisms, we show that the resulting hybrid model is capable of reproducing the social power distribution patterns observed in real-world platforms that are intuitively influenced by both meritocratic principle and Matthew-effect mechanism. This lends credence to interpreting the two models as the two extremes of a conceptual spectrum underlying social network formation. 

Beyond analyzing the structural differences across various types of online social platforms, the shift in underlying network formation mechanisms also influences user behavior patterns. Understanding these mechanisms can thus provide deeper insights into all kinds of phenomena on arising online platforms. On content-based networks, the intensified competition for social power may affect creators’ strategic decisions when generating content, thereby shedding light on the evolving dynamics of the online content ecosystem. For instance, a widening social power gap between individuals of comparable quality may encourage content creators to distinguish themselves in terms of topics, style, and other aspects, thus helping mitigate the tendency towards content homogenization. Moreover, the meritocracy-based mechanism facilitates more responsive interactions between content creators and audiences, enabling user preferences to exert a quicker and more explicit influence on the emergence of social media influencers.


\section*{Methods}

\subsection*{Convergence analysis of the meritocracy-based model}

Given the two conditions for each node to reach an equilibrium state, the probability of the network reaching equilibrium within $t$ time steps is
\begin{align*}
    \mathbb{P}[a_{12}(t)=1||d_1^{out}(t)=M,&a_{21}(t)=1||d_2^{out}(t)=M,\\..., &a_{N1}(t)=1||d_N^{out}(t)=M].
\end{align*}

Since the out-links of different nodes are independent of each other, for any $j_1,j_2 \in \{1,2,...i-1,i+1,...,N\},j_1 \neq j_2$, $a_{j_1i}(t)$ and $a_{j_2i}(t)$ as well as $d_{j_1}^{out}(t)$ and $d_{j_2}^{out}(t)$ are independent of each other. Therefore the above expression can be split into the product of $N$ probabilities:
\[
	\mathbb{P}[a_{12}(t)=1||d_1^{out}(t)=M] \cdot \prod\limits_{j=2}^{N} P[a_{j1}(t)=1||d_j^{out}(t)=M],
\]
which is larger than
\[
	\mathbb{P}[a_{12}(t)=1] \cdot \mathbb{P}[a_{21}(t)=1] \cdot \cdot \cdot \mathbb{P}[a_{N1}(t)=1].
\]

The probability that node $i\neq 1$ has not been linked to node $j$ (or node $1$ has not been linked to node $2$) within the first $t$ time steps is equal to $(\frac{N-2}{N-1})^t$. Therefore, $P[a_{12}(t)=1]=P[a_{21}(t)=1]=...=P[a_{N1}(t)=1]=1-(\frac{N-2}{N-1})^t$, which leads to the conclusion
\begin{align*}
    &\mathbb{P}[\text{An equilibrium is reached within}\enspace t \enspace \text{time steps}] \\ \geq &\left(1-\left(\frac{N-2}{N-1}\right)^t\right)^N.
\end{align*}
	
When $t\rightarrow +\infty$, $(1-(\frac{N-2}{N-1})^t)^N\rightarrow 1$, the convergence is thus guaranteed.

\subsection*{Expected in-degree approximation of the meritocracy-based model}

The expected in-degree of node $i$ in the meritocracy-based model can be precisely computed by the following explicit formula:

\[
\mathbb{E}[d_i^{in}] = P(M,i) =	\sum_{k=0}^{M-1} 
	\left ( 
	\sum_{A: A \subset \{i,i+1,...,N-1\},\ |A|=k}\  \prod_{j \in A} \frac{1}{j}
	\right ).
\]

For sufficiently large network size $N$, by using integrals to approximate sums in equation \eqref{eq:explicit indegree}, for example, in the case of $k=1$,
\[
\begin{aligned}
   \frac{1}{i} + \frac{1}{i+1} + ... + \frac{1}{N-1} &= \frac{1}{N} \left[\frac{1}{\frac{i}{N}} + ... + \frac{1}{\frac{N-1}{N}}\right] \\
	&\approx \int_\frac{i}{N}^\frac{N}{N} \frac{1}{x} \mathrm{d}x = \log\frac{N}{i}.
\end{aligned}
\]
We can obtain an approximation:
\[
    \mathbb{E}[d_i^{in}] \approx \sum_{m=0}^{M-1} \frac{1}{m!}(\log\frac{N}{i})^m, \ 1 \leq i \leq N, N \rightarrow \infty.
\]

\subsection*{Analytical results for the verification of power-law in-degree distribution}

For the meritocracy-based model, recall that when the network size $N$ is sufficiently large, equation \eqref{eq:approximate indegree} can be used as a close approximation of the expected in-degree. Denote $\sum_{m=0}^{M-1} \frac{1}{m!}\left(\log \frac{N}{i}\right)^m = H(i,M) \cdot \frac{N}{i}$, for any $k \in \mathbb{N}^*$, for any small $\epsilon$, let
\[
H(i^-,M) \cdot \frac{N}{i^-} = k,\ H(i^+,M) \cdot \frac{N}{i^+} = k+\epsilon
\]
Since $\frac{\partial H}{\partial x} \rightarrow 0,\ N \rightarrow \infty$, we have
\[
i^{-}-i^{+} \approx \frac{N(1-H_M)}{k(k+\epsilon)} \sim k^{-\alpha},\ \alpha \in (2,3),
\]
which suggests that at the equilibrium state, the in-degree distribution of nodes in the meritocracy-based model follows a power-law distribution.

For the Matthew-effect-based model, start with the approximate expected in-degree \eqref{eq:Matthew indegree}. For any in-degree $d$ and time $t$, let $i_t(d)$ be the ranking of the node with in-degree $d$ at time $t$. Then we have
\begin{align*}
    &\frac{2(M+1)N^2}{(N+i_t(d)-1)(N+i_t(d))} - 1 = d \\ &\Rightarrow i_t(d) \approx \left(\left(\frac{2(M+1)}{d+1}\right)^\frac{1}{2} - 1\right)N,
\end{align*}
which means that there is a fraction of $2 - \left(\frac{2(M+1)}{d+1}\right)^\frac{1}{2}$ nodes whose expected in-degree is larger than $d$. Therefore, we have the probability density function of in-degree 
\[
    f(d) \approx \left(\frac{M+1}{2}\right) (d+1)^{-\frac{3}{2}},
\]
which suggests that at the equilibrium state, the in-degree distribution of nodes in the Matthew-effect-based model follows a power-law distribution.

\subsection*{Bilibili and Twitch data collection}

For content-based networks, we collect real-world user data from Bilibili and Twitch. Bilibili is an online video-sharing platform, while Twitch is a live-streaming platform; both allow users to independently generate and share content. Both platforms provide built-in content categories, making it feasible to collect user data within a single category, which better satisfies the meritocracy-based framework that treats the population size as constant.

For Bilibili, as data collection can only be carried out through video information, we focus on a specific content category at one time and extract user data for those who uploaded videos in that category within the past three months. For Twitch, since data collection can only capture users who are currently live-streaming, we track a specific streaming category over the course of two weeks and collect data on users who streamed in that category during this period.

\bibliography{thesis}

\end{document}